%% file: HOMpaper.tex
\documentclass[sigconf]{acmart}
\renewcommand\footnotetextcopyrightpermission[1]{}
\setcopyright{none}
\usepackage[T1]{fontenc}
\usepackage{booktabs} 
\usepackage{mathtools}
\usepackage{microtype}
\usepackage{listings}
\usepackage{pgfplots}
\pgfplotsset{compat=1.14}
\usepackage{multirow} 
\usepackage{pgfplotstable}
\usepackage{threeparttable} 
\usepackage{url}
\usepackage{pifont}
\usepackage{paralist}
\usepackage[scaled]{beramono}
\usepackage{microtype}
\usepackage{subcaption}
\usepackage{subcaption}
\usepackage[skins]{tcolorbox}
\newcommand\searchvar{$\texttt{search}_\textit{var}$}
\newcommand\searchpri{$\texttt{search}_\textit{pri}$}
\newcommand\searchbf{$\texttt{search}_\textit{bf}$}
\newcommand\searchgen{$\texttt{search}_\textit{gen}$}

\newcommand{\sshom}{SSHOM}
\newcommand{\sshoms}{SSHOMs}

\definecolor{pgreen}{rgb}{0.25,0.5,0.37}
\newcommand{\lstbg}[3][0pt]{{\fboxsep#1\colorbox{#2}{\strut #3}}}
\lstdefinelanguage{diff}{
	basicstyle=\ttfamily,
	morecomment=[f][\lstbg{red!20}]-,
	morecomment=[f][\lstbg{green!20}]+,
}
\newlength{\MyColorBoxWidth}
\newcommand{\MyColorBox}[2][red]%
{%
    \settowidth{\MyColorBoxWidth}{#2}%
    \colorbox{#1}%
    {%
        {%
            \parbox[c][3pt][c]{\MyColorBoxWidth}{\centering#2}%
        }%
    }%
}

\definecolor{S100}{RGB}{246, 65, 45}
\definecolor{S66}{RGB}{255, 86, 7}
\definecolor{S50}{RGB}{255, 152, 0}
\definecolor{S33}{RGB}{255, 236, 25}
\definecolor{S0}{RGB}{0, 255, 0}
\newtcbox{\mybox}[2][red]{
    on line, 
    arc=3pt,
    colback=#1!50!white,
    colframe=#1!50!black,
    boxrule=0.5pt,
    boxsep=0pt,
    left=3pt,
    right=3pt,
    top=1pt,
    bottom=1pt,
    sharp corners=#2,
}
\newcommand{\dangerbox}[1]{\mybox[S100]{northwest}{#1}}
\newcommand{\suspiciousbox}[1]{\mybox[S50]{northeast}{#1}}
\newcommand{\cautionbox}[1]{\mybox[S33]{north}{#1}}
\newcommand{\safebox}[1]{\mybox[S0]{all}{#1}}

\renewcommand{\paragraph}[1]{\noindent\textbf{\emph{#1}}. } 

\begin{document}
\title{Efficiently Finding Higher-Order Mutants}



\author{Chu-Pan Wong}
\affiliation{%
 \institution{Carnegie Mellon University}
}
\author{Jens Meinicke}
\affiliation{%
 \institution{Carnegie Mellon University}
}
\author{Leo Chen}
\affiliation{%
 \institution{Carnegie Mellon University}
}
\author{Jo\~{a}o P. Diniz}
\affiliation{%
 \institution{Federal University of Minas Gerais}
}
\author{Christian K\"{a}stner}
\affiliation{%
 \institution{Carnegie Mellon University}
}
\author{Eduardo Figueiredo}
\affiliation{%
 \institution{Federal University of Minas Gerais}
}

\begin{abstract}

Higher-order mutation has the potential for improving major drawbacks of
traditional first-order mutation, such as by simulating more realistic faults
or improving test optimization techniques. Despite interest in studying
promising higher-order mutants, such mutants are difficult to find due to the
exponential search space of mutation combinations. State-of-the-art
approaches rely on genetic search, which is often incomplete and expensive
due to its stochastic nature. First, we propose a novel way of finding a
complete set of higher-order mutants by using \emph{variational execution}, a
technique that can, in many cases, explore large search spaces completely and
often efficiently. Second, we use the identified complete set of higher-order
mutants to study their characteristics. Finally, we use the identified
characteristics to design and evaluate a new search strategy, independent of
variational execution, that is highly effective at finding higher-order
mutants even in large code bases.

\end{abstract}


\maketitle

\pagestyle{empty}

\input{sections/introduction}

\input{sections/HOM}

\input{sections/approach.tex}
\input{sections/evaluation.tex}
\input{sections/algorithm.tex}

\input{sections/relatedwork}
\input{sections/conclusion}
\bibliographystyle{ACM-Reference-Format}
\bibliography{bibtex/MYfull,bibtex/literature,bibtex/refsjp,bibtex/zotero}

\appendix
\input{sections/appendix}

\end{document}

%% file: sections/introduction.tex
\section{Introduction}
\label{sec:intro}

Mutation analysis has been studied for decades in software engineering
research~\cite{papadakisMutation19}, and increasingly adopted in industry
in recent years~\cite{petrovicIndustrial18, petrovicState18}. Mutation analysis has many applications, including assessing and improving test suite quality,
generating or minimizing a test suite, or as a proxy for evaluating
other research techniques such as fault
localization~\cite{justAre14,papadakisMutation19}. Traditionally, mutation analysis
injects syntactic mutations into an existing program and runs the existing
test suite to assess whether the test suite is sensitive enough to detect the
mutations. \looseness=-1



\emph{Higher-order mutation} is the idea of combining multiple mutations with
the goal of representing more subtle changes, more complex changes, or
changes that better mirror human mistakes~\cite{jiaHigher09}. To that end,
\citet{jiaHigher09} distinguish \textit{first-order mutants}, consisting of
a single change, from \textit{higher-order mutants} that combine multiple
changes. While most research on mutation analysis has focused on first-order
mutants, higher-order mutation is promising: For example, recent studies claim that
higher-order mutants are less likely to be equivalent
mutants~\cite{mateoValidating13,kintisEvaluating10,papadakisEmpirical10,madeyskiOvercoming14},
and that higher-order mutants can reduce test
effort~\cite{jiaConstructing08,jiaHigher09,poloDecreasing09}. In
Section~\ref{sec:hom}, we discuss a specific use case of higher-order mutants with
a motivating example.




A key challenge in adopting higher-order mutation 
is \emph{identifying} beneficial higher-order mutants.
Most higher-order mutants are as easy to kill as their constituent
first-order mutants, due to coupling.
\citet{jiaHigher09} argue that only a subset of all possible combinations
better simulate real faults and increase subtlety of the seeded faults.
Specifically, \citet{jiaHigher09, jiaConstructing08} look for what
they name a \textit{strongly subsuming higher-order mutant (SSHOM)},
a particular kind of higher-order mutant which is harder to detect than its constituent first-order mutants,
as we will explain in Section~\ref{sec:hom}.
However, SSHOMs are tricky to find among the vast quantity of
possible combinations of first-order mutants.
Current approaches use genetic-search techniques,
guided by a simple fitness function~\cite{jiaHigher09,jiaConstructing08,harmanAngels14,langdonEfficient10}.
Since SSHOMs are difficult to find, little is known about them and their characteristics.

In this work, we develop a technique that can find a \emph{complete} set of
SSHOMs for \emph{small to medium-sized} programs, which enables us to study characteristics
of SSHOMs. Based on the identified characteristics, we then develop a new heuristic search
technique that is lightweight, scalable, and practical.  Overall,
we proceed in three steps: \looseness=-1

\textbf{(1) Variational Search:}
For the purpose of studying SSHOM in a controlled setting, we develop a new search strategy \searchvar{} that allows us to find a \textit{complete}
set of higher-order mutants for a given test suite and given set of first-order
mutants in small to medium-sized programs.
Specifically, we use
\emph{variational execution}~\cite{meinickeEssential16,wongFaster18,nguyenExploring14}, a dynamic-analysis
technique that jointly explores many similar executions of a program.
Conceptually, our approach searches for all possible higher-order mutants \emph{at
the same time}, identifying, with a propositional formula for each test case, which mutants
and combinations of mutants cause a test to fail.
From these formulas, we then encode search as a \emph{Boolean satisfiability problem}
and use BDDs or SAT solvers to enumerate
\textit{all} SSHOMs.
A complete exploration with variational execution is often feasible
for \emph{small to medium-sized} programs,
because
variational execution shares commonalities among repetitive executions
and 
because modern SAT solving techniques are relatively fast.
Though it does not scale to all programs, analyzing a complete set of SSHOMs for smaller 
programs allows us to study SSHOMs more systematically.

\textbf{(2) Complete-Mutant-Set Analysis:}
We study the characteristics of the identified higher-order mutants from Step~1.
Where previous approaches found only few samples of higher-order mutants,
we have a unique opportunity to study the characteristics of
higher-order mutants on a \textit{complete} set.
We analyze characteristics, such as, the typical number of mutants combined
and their distance in the code. This helps us better
understand higher-order mutants without the potential sampling bias from a
search heuristic. For example, we found that most SSHOMs are composed of fewer than
4 first-order mutants and that constituent first-order mutants tend to locate within
the same method or the same class.

\textbf{(3) Prioritized Heuristic Search:}
Finally, we develop a second new
search strategy \searchpri{} that prioritizes likely promising combinations of first-order
mutants based on the characteristics identified in Step~2.
The \searchpri{} is easy to implement and does not require
the heavyweight variational analysis of \searchvar{}.
Although it no longer provides completeness guarantees, it is highly efficient
at finding higher-order mutants fast and scales much better to larger
systems with tens of thousands of first-order mutants. We evaluate the new
search strategy using a different set of larger systems to avoid potential overfitting.
Our results indicate that the previously identified characteristics are useful in
guiding the search. Our new search strategy can find a large number of SSHOMs
despite an exponentially large search space, whereas existing search approaches can
barely find any.\looseness=-1




We make the following contributions in this work:
\begin{compactitem}

	\item We propose a novel way of using variational execution to find a
		\emph{complete} set of SSHOMs for small to medium-sized programs, by
		formalizing the search as a Boolean satisfiability problem.  An
		evaluation shows that we can achieve completeness and simultaneously
		increase efficiency. (Section~\ref{approach}) \looseness=-1


	\item Using a complete set of SSHOMs, we make a first step in studying
		basic characteristics of SSHOMs with the goal to inform future
		research. (Section~\ref{sec:evaluation})

	\item To show how useful the characteristics are, we use them to design a
		new lightweight prioritized search strategy, independent of variational execution.
		We evaluate the prioritized search strategy on a fresh set of larger
		benchmarks, showing that the new search is scalable and generalizable.
		(Section~\ref{sec:algorithm})

\end{compactitem}

%% file: sections/HOM.tex
\section{Higher-Order Mutants}
\label{sec:hom}

\newcommand\Pass{\ding{51}}
\newcommand\Fail{\ding{55}}
\begin{figure}[t]
   \centering
   \includegraphics[width=\linewidth]{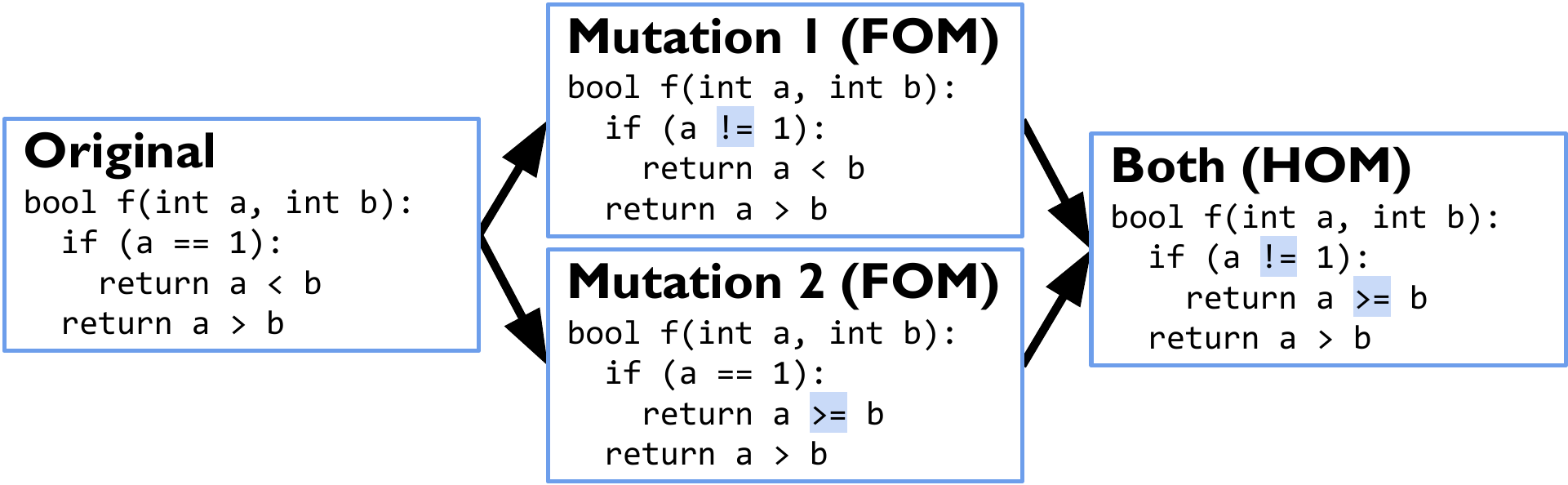}
   \footnotesize
   \vskip 1em
   \begin{tabular}{llllll}\toprule
    Test&
Orig.&
Mut.\ 1&
Mut.\ 2&
Both&
Failure Cond.\\\midrule
$T_1$: \lstinline.assert f(1, 2).&
\Pass&
\Fail&
\Fail&
\Fail&
$m_1\vee m_2$\\
$T_2$: \lstinline.assert !f(0, 3).&
\Pass&
\Fail&
\Pass&
\Pass&
$m_1\wedge\neg m_2$\\
$T_3$: \lstinline.assert !f(1, 1).&
\Pass&
\Pass&
\Fail&
\Pass&
$\neg m_1\wedge m_2$\\\bottomrule
   \end{tabular}
   \caption{Example of mutations with their test outcomes.}
   \Description{Example with two mutations and corresponding test outcomes.}
   \label{fig:hom}
   \vspace{-2em}
\end{figure}


Mutation analysis introduces a set of
syntactic changes to a software artifact
and observe whether the previously passing test suite
is sensitive
enough to detect the changes (termed ``to kill the mutant'').
Traditionally, many simple small changes are explored
in isolation, one at a time; several catalogues
of mutation operators 
that perform small syntactic changes exist~\cite{king1991fortran, papadakisMutation19}.

In its simplest form, \textit{higher-order mutants} are combinations
of two or more first-order mutants~\cite{jiaHigher09,jiaConstructing08}.
The set of possible second-order mutants grows quadratically with
the size of the set of first-order mutants from which they are combined;
if considering combining more than two first-order mutants, the
set of possible higher-order mutants grows much faster.

Many higher-order mutants are of little
value in practice, because a test that would kill any
constituent first-order mutant will likely also kill the higher-order mutant,
discussed as the coupling effect hypothesis~\cite{offuttInvestigations92}.
However, \citet{jiaHigher09} show that there exist several classes of
higher-order mutants that are potentially valuable, because they
exhibit interesting behavior. They specifically highlight
\textit{strongly subsuming higher-order mutant (SSHOM)}, 
in which the constituent mutants interact in ways making the
higher-order mutant hard to kill, as we will explain in
detail in Section~\ref{sec:sshom}.

\input{sections/usecases.tex}

\subsection{Strongly Subsuming Higher-Order Mutants (SSHOMs)}\label{sec:sshom}



\citet{jiaHigher09} classify higher-order mutants into several kinds,
specifically highlighting SSHOMs as useful. For this reason, our
work targets SSHOMs, though we expect that it can be generalized to other classes of higher-order
mutants. Specifically, \citet{jiaHigher09} define a SSHOM as a higher-order
mutant that can only be killed by a subset of test cases that kill all its
constituent first order mutants. More formally, let $h$ be a higher order
mutant composed of first-order mutants $f_1, f_2, \ldots, f_n$, $T_h$ the set
of test cases that kill the higher-order mutant $h$, and $T_i$ the set of
test cases that kill the first-order mutant $f_i$, then $h$ is a SSHOM if and
only if:

\begin{equation}\label{eq:sshom}
T_h \neq \emptyset \quad\wedge\quad T_h \subseteq \bigcap_{i\in 1\ldots n} T_i
\end{equation}

If we further restrict $T_h$ to be a strict subset, we get a even stronger type of SSHOM, which we
call \emph{strict} strongly subsuming higher order mutant, denoted as \textit{strict-SSHOM}.\footnote{SSHOMs have been
defined inconsistently in the literature as subset~\cite{harmanAngels14} and strict subset~\cite{jiaHigher09, jiaConstructing08}.
We inherit the definition of SSHOMs from \citet{harmanAngels14}, as it is the most recent work. As we will see in the evaluation,
the difference between subset and strict subset is significant, so we make the distinction explicit, introducing strict-SSHOM as a distinct subclass and reporting results on both.}
In other words, there must be at least one test case that kills a
first-order mutant, but not the higher-order mutant.
Thus, in a strict-SSHOM, multiple first-order mutants interact such
that they mask each other at least for some test cases, making the strict-SSHOM
harder to kill than all the constituent first-order mutants together.


Our (manually constructed) SSHOM in Figure~\ref{fig:hom} illustrates this relation:
Intuitively, the first first-order mutant (replacing `\texttt{==}' by `\texttt{!=}') forces the execution to go into an unexpected branch, and
the second (replacing `\texttt{<}' by `\texttt{>=}') inverts the return values. The two changes in control and data flow are easy to detect separately (i.e., killed by two test cases each),
but the combination of them is more subtle and only detected by one test case.

\subsection{Finding SSHOMs}

An SSHOM is defined in terms of subset relation among mutants killed by a set
of test cases. For a given set of first-order mutants, the search space is
finite, though very large due to the combinatorial explosion. Since only few
of the combinations are interesting and those are hard to find in vast search
spaces, higher-order mutation testing has long been considered too expensive.

\citet{jiaHigher09} explored search techniques to find SSHOMs, finding that
genetic search performs best.
We will use their genetic-search strategy, together with a brute-force strategy, as baselines for our evaluations.
Although genetic search has been shown to successfully find
SSHOMs, it requires considerable resources to evaluate
many candidates, involves significant
randomness, and cannot give guarantees of completeness, e.g., establish that
no SSHOM exists or enumerate them all.

All existing techniques for finding higher-order mutants (including this
work) require executing all first-order mutants with a fixed test suite as
part of constructing higher-order mutants, as we will discuss in
Section~\ref{approach}. As such, our work is less appealing to 
the traditional mutation testing use case of
evaluating test suite adequacy. However, we
argue that finding SSHOMs is still valuable as a research tool for fault
localization and program repair, as discussed in
Section~\ref{subsec:usefulness}. In line with our work, dominant mutants have
been motivated and investigated as a research tool to improve mutation
testing research, which we discuss in Section~\ref{sec:relatedwork}.

%% file: sections/usecases.tex
\subsection{Usefulness of Higher-Order Mutants}
\label{subsec:usefulness}

A recent survey of over 39 papers on higher-order mutation testing~\cite{ghidukHigher17} summarized a large number of different application scenarios for higher-order mutants claimed in prior research, including  mutant
reduction~\cite{harmanAngels14,iidaReducing17,ghidukReducing16}, coupling effect analysis~\cite{gopinathTheory17,jiaHigher09}, equivalent
mutant reduction~\cite{kintisEmploying15,madeyskiOvercoming14}, test data evaluation~\cite{harmanStrong11}, and test suite
reduction~\cite{harmanAngels14,mateoValidating13}. 
In the following, we illustrate a concrete example of how higher-order mutations can be useful to software-engineering researchers for creating synthetic, but challenging faults to evaluate various software engineering tools.

The effectiveness of many approaches in software-engineering research
needs to be evaluated on faults in software systems.
For example, fault localization tools need to evaluate how accurately
they can localize the faults, test suite generation tools need to
evaluate how effective the generated tests are at finding bugs, and
program repair tools need to evaluate how many faults they can repair.
When evaluating their tools, researchers often have the choice of running
evaluations on a curated,
often small, set of real bugs or running
on large numbers of synthetically seeded bugs.
Both approaches have known benefits and drawbacks:
\begin{compactitem}
\item Seeded faults are convenient: Easy to create and providing a perfect
ground truth, they allow researchers to run experiments with 
very large numbers of faults on almost any system.
For example, fault localization techniques were often evaluated on 
artificially seeded
single-edit faults, such as those in the \textit{Siemens} test
suite~\cite{hutchins1994experiments} (e.g.,
\cite{pearson2017evaluating,jones2005empirical,renieres2003fault,liu2005sober,abreu2007accuracy}).
Researchers have been critical of this style of evaluation, arguing
that seeded single-edit faults are not representative of most
real faults (which often require fixes in multiple locations)~\cite{justAre14,
zhongEmpirical15} and that that fault localization techniques may not
generalize as they are over-optimized
in finding such simple single-edit faults~\cite{pearson2017evaluating}.

\item In contrast, if curated well, datasets of real faults can be much more
representative of realistic usage scenarios.
Research on automated program repair is almost exclusively evaluated
on a few dozen to a few hundred real faults~\cite{le2013current}.
For example, the widely used \textit{Defects4J} dataset~\cite{justDefects4J14}
curated 438~faults with corresponding failing tests from
5~libraries.
Creating high-quality datasets of realistic and representative faults
is challenging and typically requires significant human and engineering
effort~\cite{Madeiral2019,justDefects4J14,madeiralBears19,tomassiBugSwarm19}.
Therefore, while it is easy to seed millions of faults in almost any
program, only few datasets of curated real faults are available, 
often only with 
moderate numbers of faults in a small number of libraries or programs.
Some researchers warn that overly focusing on few datasets of faults,
such as Defects4J, 
leads to repair approaches that often overfit the available faults~\cite{durieux2019empirical,tomassiBugSwarm19}.
\end{compactitem}

In this tension between simple seeded faults and expensive to curate real faults,
higher-order mutation may provide a compromise.
Certain kinds of higher-order mutants, in particular
SSHOMs that we study in this work, are more \emph{subtle} and 
\emph{hard to kill} (shown both theoretically~\cite{gopinathTheory17} and
empirically~\cite{langdonEfficient10, jiaConstructing08, jiaHigher09,
harmanManifesto10, omarSubtle17}). They are more promising
to simulate real faults than traditional first-order mutants:
For example,
\citet{zhongEmpirical15} and \citet{justAre14} found that more than 70\,\%,
respectively 50\,\%, of real faults are caused by faults in more than two
locations. \citet{justAre14} also found that $73\%$ of real faults are
coupled to mutants, while on average 2 mutants are coupled to a single real
fault. 
That is, certain kinds of higher-order mutants may be more representative
of real faults. Thus, assuming we can find them efficiently, which is the goal of this paper, we can still automate their creation and seed
thousands of these more challenging faults in almost any software systems.

\begin{figure}
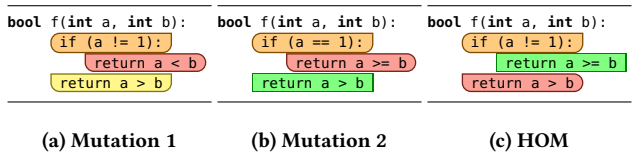

    \begin{subfigure}{0.32\columnwidth}
    \centering
    \begin{lstlisting}[language=python, frame=tb, tabsize=2, basicstyle=\ttfamily\scriptsize, escapechar=~]
bool f(int a, int b):
     ~\suspiciousbox{if (a != 1):}~
         ~\dangerbox{return a < b}~
     ~\cautionbox{return a > b}~
    \end{lstlisting}
    \caption{Mutation 1}
    \end{subfigure}
    \begin{subfigure}{0.32\columnwidth}
    \begin{lstlisting}[language=python, frame=tb, tabsize=2, basicstyle=\ttfamily\scriptsize, escapechar=~]
bool f(int a, int b):
    ~\suspiciousbox{if (a == 1):}~
        ~\dangerbox{return a >= b}~
    ~\safebox{return a > b}~
    \end{lstlisting}
    \caption{Mutation 2}
    \end{subfigure}
    \begin{subfigure}{0.32\columnwidth}
    \begin{lstlisting}[language=python, frame=tb, tabsize=2, basicstyle=\ttfamily\scriptsize, escapechar=~]
bool f(int a, int b):
    ~\suspiciousbox{if (a != 1):}~
        ~\safebox{return a >= b}~
    ~\dangerbox{return a > b}~
    \end{lstlisting}
    \caption{HOM}
    \end{subfigure}
\caption{Suspicious lines based on coverage ranking using spectrum-based fault localization~\cite{jones2005empirical}. Ranking is shown as intensity of \dangerbox{danger}, \suspiciousbox{suspicious}, \cautionbox{caution} and \safebox{safe}.}
\Description{Suspicious lines based on coverage ranking using spectrum-based fault localization~\cite{jones2005empirical}. Ranking is shown as intensity from red (danger), orange (suspicious) over yellow (caution) to green (safe).}
    \label{fig:tarantula}
\end{figure}

Let us illustrate the potential of higher-order mutation for 
fault localization with a small example with 3 existing test cases in Figure~\ref{fig:hom}. The program is
mutated into two first-order mutants, which are later combined to form a
higher-order mutant, with test results for each mutant reported in the figure. 
Note how this higher-order mutant fails for fewer test cases than the 
constituent first-order mutants.
In this simple setting, the classic fault localization technique Tarantula~\cite{jones2005empirical} works quite well for the 
first-order mutants, highlighting the mutated lines as shown in Figure~\ref{fig:tarantula}; but for the higher-order mutant, Tarantula
fails to report the two mutated lines, but instead marking the unchanged line
as dangerous.  This example shows how fault localization fails to locate the
faulty lines if the mutations are interacting with each other, which, 
as discussed, may be
expected for realistic faults~\cite{justAre14, zhongEmpirical15}. 
As a further consequence, a
program repair technique based on spectrum-based fault localization may not
even attempt to fix the first return statement~\cite{le2012systematic}.
\looseness=-1

To realize the full potential of higher order mutants
for these and other use cases, it is critical to have
an efficient way of finding interesting higher-order mutants. In this work, we do not reevaluate the usefulness
of HOMs for various use cases~\cite{ghidukHigher17} or how well they represent real
faults~\cite{justAre14,zhongEmpirical15,jiaHigher09}, which has been studied
repeatedly and comprehensively in prior work~\cite{do2019systematic}.
Instead, we focus on a technical problem that made SSHOMs too costly and
impractical: \emph{How to efficiently find SSHOMs} (and for part of our
research also how to find all SSHOMs in small to medium-sized programs so
that we can study their characteristics).

%% file: sections/approach.tex
\section[Step 1: Complete Search With Variational Execution]{Step 1: Complete Search With Variational~Execution (\searchvar)}
\label{approach}



\label{subsec:overview}


In this step, we develop \searchvar{} to compute a \emph{complete} set of
SSHOMs so that we can study the properties of SSHOMs.  

First, given a program under analysis, we generate all first-order mutants upfront by
applying our mutation operators exhaustively at every applicable location.
We represent each mutant as a Boolean option and use a ternary
conditional operator to encode the change. 
For example, in the code snippet below, we show how we encode the two first-order mutants from
Figure~\ref{fig:hom}.

\noindent\begin{minipage}{\linewidth}
\begin{lstlisting}[language=python, basicstyle=\ttfamily\scriptsize, frame=tb, tabsize=2, escapechar=~]
bool f(int a, int b):
    if (~\colorbox{gray}{m1 ? a != 1 : a == 1}~):
        return ~\colorbox{gray}{m2 ? a >= b : a < b}~
    return a > b
\end{lstlisting}
\end{minipage}



After encoding first-order mutants, we use variational execution as a black-box technique to explore which test cases
fail under which combinations of first-order mutants. In a nutshell, variational execution runs the program under
analysis by dynamically tracking the differences caused by options 
(similar to executing the program symbolically with symbolic values for all mutations)~\cite{meinickeEssential16,nguyenExploring14,wongFaster18,austinMultiple12}. Conceptually,
a single run of variational execution with options is equivalent to running all combinations of options sequentially,
but it is usually much faster due to sharing of similar executions at runtime~\cite{meinickeEssential16,nguyenExploring14,wongFaster18}.
For a given test execution, variational execution will return a propositional formula
representing exactly the combinations of options for which the test fails,
which we illustrate for our running example in Figure~\ref{fig:hom} (last table column).\looseness=-1





Finally, we collect all propositional failing conditions for all test cases and
use them to search for SSHOMs by encoding the search as a Boolean satisfiability problem.
Using BDDs or SAT solvers, we can then enumerate all solutions,
which correspond directly to all SSHOMs.
Although the formulas can be large if we have many first-order mutants and
test cases and finding satisfiable assignments is NP-hard,
modern SAT solving techniques are scalable enough 
with small to medium-sized systems.
Our implementation is available online.~\footnote{\url{https://figshare.com/s/182142e4e7dc3b5981ff}}

\subsection{Mutant Generation}



We represent each first-order mutant with
a Boolean option (global static field in Java) and encode the pending change
with a ternary conditional operator. We encode all first-order mutants all
at once into the program to generate a metaprogram, which is used in our later
steps for finding SSHOMs. This compact encoding of mutants defines a finite set
of first-order mutants, which is critical for variational execution to be
efficient~\cite{wongTesting18}. Similar encodings have been explored in the
past in different contexts, such as speeding up mutation
testing~\cite{untch1993mutation,madeyski2010judy,justUsing11}. Using this
encoding, we also ensure a fair comparison with baseline approaches by
excluding compilation time and using the same metaprograms. \looseness=-1



For our experiments, we implemented 3~mutation operators:
(1)~\textit{Arithmetic Operator Replacement} (AOR, mutating \texttt{+}, \texttt{-}, \texttt{*}, \texttt{/}, \texttt{\%})
(2)~\textit{Relational Operator Replacement} (ROR, mutating \texttt{==}, \texttt{!=}, \texttt{<}, \texttt{>}, \texttt{<=}, \textit{>=})
and (3)~\textit{Logical Connector Replacement} (LCR, mutating \texttt{||} and \texttt{\&\&}).
These comprise three of five most useful mutation operators according to
Offutt et al.~\cite{offuttExperimental96, offuttExperimental93},
excluding two further based on recent insights:
(4)~\textit{Absolute Value Insertion} (ABS) has been shown to be less useful in
practice~\cite{petrovic2018state}, so we excluded it to avoid a meaninglessly large search space.
(5)~\textit{Unary Operator Insertion} (UOI) would add many more  mutants, most of which are likely equivalent to the ones generated
from other mutation operators (e.g., mutating \texttt{a+b} to \texttt{a+-b} using UOI is
equivalent to \texttt{a-b} using AOR)~\cite{king1991fortran,madeyski2010judy, petrovic2018state}.

We argue that our selection of mutation operators is sufficient as the first
step in studying properties of SSHOMs. A recent study by
\citet{kurtzAnalyzing16} suggests that mutation operators should be carefully
chosen for individual programs to maximize the benefits of mutation testing,
while our mutation operators remain a reasonable choice across programs. With
the goal to uncover general characteristics of SSHOMs, we decided to use
this small but well studied set of mutation operators.

We generate all possible mutations \textit{exhaustively} at every applicable location in the source code.
To apply multiple mutation operators to the same expression, we nest ternary
conditional operators. \looseness=-1

\subsection{Variational Execution}\label{subsec:ve}

We use variational execution to determine which combinations of mutants
fail a test case. The novelty of using variational execution lies in the efficient and complete
exploration of all mutants, as opposed to one mutant at a time in traditional search-based approaches.
For this work, the details of how variational execution works
are not relevant, and we use it as a black-box technique.
Here, we only provide an intuition and refer the interested readers to existing
literature for a more in-depth discussion~\cite{austinMultiple12,nguyenExploring14,meinickeEssential16,wongFaster18}.

Variational execution performs computations with \emph{conditional values}~\cite{wongFaster18},
which may represent multiple alternative concrete values. For example, a conditional value
\texttt{<$\alpha$, 1, -1>} indicates that \texttt{x} has the
value \texttt{1} if $\alpha$, and \texttt{-1} otherwise; conditional values
can represent a \emph{finite} number of alternative \emph{concrete} values distinguished
by propositional conditions over symbolic options.
Variational execution then computes with conditional values and propagates them
along data and control flow, possibly under symbolic path conditions.
In a nutshell, variational execution can be considered
as an extreme design choice among
various forms of symbolic program evaluation~\cite{K76,biere1999symbolic,ALM:SE08,senMultiSE15,bornholt2018finding}
for finite domains,
in which computations are maximally performed
on concrete values, but boolean symbolic
values may distinguish between multiple
concrete values per variables~\cite{meinickeEssential16,wongFaster18,austinMultiple12}.


For our purposes, we consider all variables representing first-order mutants as symbolic
options (technically as a conditional value \texttt{<$m_i$, true, false>}).
This way, all state changes caused by mutants can
be compactly tracked, which enables us to explore all combinations of mutants at the same time.
As output, we determine under which combinations of mutants a test case fails
(propositional formula over first-order mutants as illustrated in Fig.~\ref{fig:hom}),
by simply observing under which condition
any asserted expression evaluates to \emph{false}. \looseness=-1

In theory, mutant interactions can
cause a combinatorial explosion in conditional values where
an exponentially many alternative values
for different combinations of mutants
need to be tracked for a single variable.
However, in practice not all mutants affect each test and not all mutants interact,
enabling often reasonably efficient exploration of all feasible combinations.
We defer the discussion of this scalability issue to Section~\ref{subsec:limitations}.


Multiple different implementations of variational execution exist for a number of
programming
languages~\cite{austinMultiple12,austinFaceted13,meinickeEssential16,nguyenExploring14,schmitzFaceted16,wongFaster18,KKB:ISSRE12}. 
We use \textit{VarexC},
a state-of-the-art implementation of variational execution for Java, based on bytecode transformation~\cite{wongFaster18}.
For this work,
we extended \textit{VarexC} to deal with infinite loops that are caused by some mutations.
Existing mutation-testing techniques often detect infinite loops by setting timeout on test
cases, but generalizes poorly to
approaches that explore multiple branches
and track alternative values.
Instead, we count how many basic blocks have been executed and terminate execution
in path conditions where a threshold  is reached
(10~million in our experiments, based on  observations of the test cases).

\subsection{ SSHOMs Search as a Boolean~Satisfiability Problem}\label{subsec:sat}


We use the output of variational execution---propositional formulas indicating under which combinations of mutations
each test fails---to construct a single formula that is satisfiable exactly for
those assignments that represent SSHOMs, based on our definition of
SSHOM in Section~\ref{sec:hom}.
This way, the search for SSHOMs is transformed into a Boolean satisfiability problem, which we can solve with BDDs or SAT solvers.
To derive the formula, we outline the criteria for identifying \sshoms{} as defined by \citet{jiaHigher09} (see Sec.~\ref{sec:sshom}) and construct a logical expression for each criterion.\looseness=-1

Let $T$ be the set of all tests, $M$ be the set of all first order mutants, and $f_t$ be the propositional formula over literals from $M$ describing the mutant configurations in which test $t\in T$ fails ($f$ is generated with
variational execution, see above). As shorthand, let $\Gamma(m,t)$ be the result of evaluating $f_t$ with first-order mutant $m$ assigned to \emph{true} and all other mutants assigned to \emph{false}; in other words, whether test $t$ fails for first-order mutant $m$. To identify \sshoms{}, we encode three criteria:\looseness=-1

\begin{enumerate}
	\item The \sshom{} must fail at least one test (i.e., must not be an equivalent mutant):
\end{enumerate}
\begin{equation}
\label{not-eq}
\bigvee\limits_{t \in T} f_t
\end{equation}

\noindent{}This check ensures that a mutant combination is killed by at least one test, encoding $T_h \neq \emptyset$ in Formula~\ref{eq:sshom} (Sec.~\ref{sec:sshom}).

\begin{enumerate}
	\setcounter{enumi}{1}
	\item Every test that fails the \sshom{} must fail each constituent first order mutant:
\end{enumerate}
\begin{equation}
\label{homsubset}
\bigwedge\limits_{t \in T} (f_t \Rightarrow \bigwedge\limits_{m \in M} 
( \neg m \vee \Gamma(m, t) )
\end{equation}

\noindent{}If a given mutant combination (i.e., higher-order mutant) is killed by a test $t$,
the same test must kill each constituent first-order mutant.
That is, for all tests and first-order mutants, the first-order mutant must either be killed by the test ($\Gamma(m,t)$) or not be part of
the higher-order mutant ($\neg m$).
This is the encoding of $T_h \subseteq \bigcap_{i\in 1\ldots n} T_i$ in Equation~\ref{eq:sshom} (Sec.~\ref{sec:sshom}).

In addition, we can optimize for \sshoms{} that are harder to kill than the constituent first order mutants, excluding those that are equally difficult to kill~\cite{jiaHigher09}.
As discussed in Section~\ref{sec:sshom}, we call these \emph{
strict-SSHOM} and require a strict subset relation in
Equation~\ref{eq:sshom} (i.e., $T_h \subset \bigcap_{i\in 1\ldots n} T_i$
rather than $T_h \subseteq \bigcap_{i\in 1\ldots n} T_i$), which requires the additional encoded condition:

\begin{enumerate}
	\setcounter{enumi}{2}
    \item There exists a test that can kill all constituent first-order mutants
    but cannot kill the strict-SSHOM.
\end{enumerate}
\begin{equation}
\label{strict-formula}
\bigvee\limits_{t \in T} \big(
\neg f_t
\wedge \bigwedge\limits_{m \in M}
( \neg m \vee \Gamma(m, t) )
\big)
\end{equation}


To find \sshoms\ and strict-SSHOMs, we take the conjunction of
Equations~\ref{not-eq}--\ref{homsubset} and \ref{not-eq}--\ref{strict-formula},
respectively, and use  BDD or SAT solver to iterate over all possible solutions.
For example, if our approach returns a satisfiable assignment in which $m_1$ and $m_3$
are selected and all other mutants are deselected, then the combination of $m_1$ and $m_3$
is a valid (strict-)SSHOM.

We use BDDs to get satisfiable solutions by default, as \textit{VarexC} uses BDDs
internally to represent propositional formulas. While constructing BDDs can be
expensive during variational execution, getting a solution from a BDD is
$\mathcal{O}(n)$, where $n$ represents the number of Boolean
variables~\cite{bryantGraphBased86}. In some rare cases where we cannot compute
a BDD due to issues like insufficient memory, we fall back to using a SAT
solver. With a SAT solver, we ask for one possible solution, then add the negation
of that solution as an additional constraint before asking for the next
solution, repeating the process until all solutions are enumerated.
We can usually
efficiently enumerate \emph{all}
possible \sshoms{} for the given set of first-order mutants and the variational-execution result of a given test suite.
\looseness=-1

\subsection{Limitations}\label{subsec:limitations}

While variational execution and the
SAT encoding provide a new strategy
to enumerate \emph{all} SSHOMs,
this approach comes also with
severe restrictions, mostly 
regarding scalability and engineering
limitations inherited from the tools we
use, which limits broad applicability in practice (which we address with an 
alternative strategy in Sec.~\ref{sec:algorithm}).


\paragraph{Combinatorial Explosion} Recent studies show that combinatorial
explosion is uncommon for the types of highly-configurable programs analyzed with variational execution in the
past~\cite{meinickeEssential16,reisnerUsing10}, mainly because programs are usually written by human
developers to have manageable interactions among options.
When applied to higher-order mutation testing, we did observe some combinatorial explosion caused by
random combinations of first-order mutants. For example, we observed cases where interactions of
first-order mutants create more than $15,000$ alternative concrete values in one single local variable.
We argue that this is the essential complexity of the mutated program, and it would be equally difficult
for other approaches to exhaustively explore a complex search space like this.
In fact, a recent similar approach that uses SMT solver to detect equivalent
mutants has similar scalability issues~\cite{kushigianMedusa19}.
However, it is possible to find efficient
search strategies when giving up the
completeness goal, as we will show in Sec.~\ref{sec:algorithm}.\looseness=-1


In the evaluation of \searchvar{}, we manually removed
some problematic first-order mutants and test cases that caused excessive
number of interactions that exceeded
our memory limits (12GB).
For fairness, we remove these mutants and test cases across all compared approaches.


\paragraph{Environment Barrier}
As other forms of symbolic evaluation, variational execution needs to deal with the
environment barrier carefully when
execution interacts with an external runtime environment that is not aware of conditional values or variability
contexts. This barrier often manifests as I/O or native method calls.
There are
several common strategies to mitigate this issue, such as creating models
for these operations~\cite{RARF:JPF11,senMultiSE15,ALM:SE08,wongFaster18}.
In our study, only few  test cases and mutants triggered
problematic environment interactions.
While solvable with more engineering effort,
we consider them noncritical for our goal
and removed the problematic tests or mutants
after manual inspection.


%% file: sections/evaluation.tex

\subsection{Evaluation}\label{subsec:oldEval}

In addition to using \searchvar{} to get a complete set of SSHOMs, we compare
efficiency and effectiveness of \searchvar{} against the existing
state-of-the-art \emph{genetic search} (\searchgen) and a baseline
\emph{brute-force} strategy (\searchbf), based on subject systems previously
used in evaluating the genetic search strategy~\cite{harmanAngels14}.












\definecolor{Gray}{gray}{0.85}
\newcolumntype{g}{>{\columncolor{Gray}}r}

\begin{table*}[t]
\centering
    \caption{Subjects and Found (strict-)SSHOMs; the last three subjects and the \textit{priority} strategy are discussed in Section~\ref{sec:algorithm}.}
    \begin{tabular}{lrr@{\ }lr@{\ }lrrrrrrrr}
        \toprule
         &  &  &&& & \multicolumn{4}{c}{Found SSHOM}  & \multicolumn{4}{c}{Found strict-SSHOM}\\\cmidrule(r){7-10}\cmidrule(l){11-14}
        Subject & LOC & Tests &(\%used)& FOMs &(\%used) & Var & Gen & BF & Pri & Var  &  Gen & BF & Pri \\\midrule
        \input{resources/stats.tex}
        \bottomrule
    \end{tabular}
    \\[.4em]
    {\footnotesize
    LOC represents lines of code, excluding test code, measured with \textit{sloccount}. Tests and FOMs
    report the numbers of test cases and first-order mutants we used in experiments, 
    with the percentages relative to the total numbers in parentheses. Var, Gen, BF, Pri denote our
    approach (Step~1, \searchvar), the genetic algorithm (\searchgen), brute force (\searchbf), and our prioritized search (Step~3, \searchpri) respectively.}

    {\footnotesize
    $\dagger$ incomplete results, solutions found with SAT solving within the 12 hours budget.}

    \label{tab:statistics}
\end{table*}

\paragraph{Subject Systems}
We replicate the setup of the most rigorous and largest previous
study on higher-order
mutation testing~\cite{harmanAngels14}.
While we cannot perform an exact replication, since we could not obtain the original
tools from the authors, not all relevant details
and parameters have been published, and some engineering limitations discussed earlier,
we still select the same subject systems
and reimplement mutation operators and search strategies in our own infrastructure.
That is, our results cannot be compared directly against the numbers reported in prior work~\cite{harmanAngels14},
but we report comparable numbers within a consistent setup.

We use the same four small to medium-sized Java programs, \emph{Monopoly},
\emph{Cli}, \emph{Chess}, and \emph{Validator}, all of which come with good
quality test suites that are deemed complete by
developers~\cite{harmanAngels14}. In addition, we use
the \textit{triangle} program commonly used in mutation
testing~\cite{jiaHigher09}. Statistics of our subject systems are shown in
Table~\ref{tab:statistics} (top).  In each subject system, we applied all our
mutation operators in all feasible locations, yielding the reported number of
first-order mutants; as discussed in Section~\ref{subsec:limitations}, we had to
exclude some mutants and test cases due to engineering limitations.



\paragraph{Baseline Search Strategies}
We compare our approach against the state-of-the-art genetic 
algorithm~\cite{jiaHigher09,harmanAngels14,jiaConstructing08} and a naive brute-force search. The
brute-force search iterates over all possible combinations of first-order mutants, starting
from all pairs, then all triples, and so on until a time limit is reached.
The brute-force search serves as a reliable baseline as there is no randomness involved
and the search is easy to implement.

We reimplemented the genetic algorithm approach based on the description in Jia et al.'s work~\cite{jiaConstructing08, jiaHigher09, jiaHigher}.
As the exact setup was not available or documented,
we leave undocumented parameters at default values.
The core of the genetic algorithm is a fitness function for
candidate higher-order mutants. Following existing work~\cite{jiaHigher09, jiaHigher, jiaConstructing08} and
using the notations in Equation~\ref{eq:sshom}, we calculate the fitness
as $\frac{|T_h|}{|\bigcap_{i\in 1\ldots n} T_i|}$.\footnote{The fitness function
has been defined either using intersect of $T_i$~\cite{jiaHigher} or
union~\cite{jiaHigher09, jiaConstructing08}. We use the former in
our reimplementation as it more precisely captures our intuition of SSHOMs.} The intuition
is that a SSHOM should fail only for a subset of test cases that kill all its constituent
first-order mutants. Thus, we use it as a
piece-wise function: a fitness of $(0, 1]$ indicates a SSHOM and $(0, 1)$ a strict-SSHOM, with lower
fitness more preferable; a fitness of 0 and larger than 1 indicate potential
equivalent mutants and non-SSHOMs, respectively, which are discarded between
generations of the genetic algorithm.

\paragraph{Measurements} All experiments were performed on AWS EC2 instances, each of which has an Intel 4-core Xeon CPU
with 16GB of RAM. We ran benchmarks to confirm that the performance is stable enough
for our measurements across different instances (especially given that we
often demonstrate order-of-magnitude differences in outcomes, which are unlikely to
stem from measurement noise).
For each search strategy (i.e., genetic algorithm, brute force, and our variational execution approach),
we measure each subject system three times and report the average, like the three restarts in the work of \citet{harmanAngels14}.
We ran each trial of genetic algorithm and brute force
for 12 hours.


\newcommand\PlotW{3.55cm}
\begin{figure}[t]
\centering
\begin{tabular}{rcc}
& SSHOM & Strict-SSHOM \\
\rotatebox[origin=c]{90}{Validator$^\dagger$} & \raisebox{-.5\height}{\includegraphics[width=\PlotW]{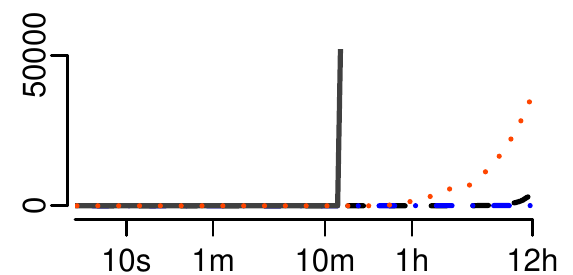}} & \raisebox{-.5\height}{\includegraphics[width=\PlotW]{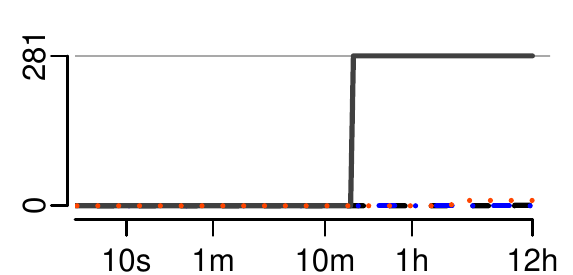}}\\
\rotatebox[origin=c]{90}{Chess$^\ddagger$} & \raisebox{-.5\height}{\includegraphics[width=\PlotW]{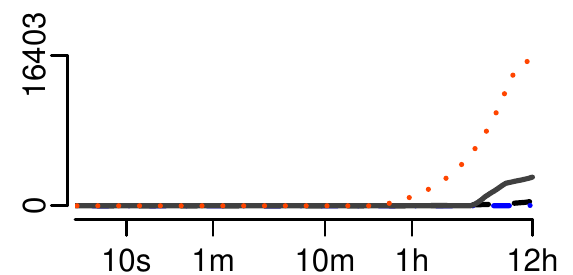}} & \raisebox{-.5\height}{\includegraphics[width=\PlotW]{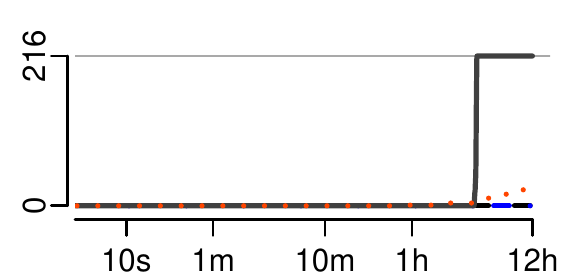}}\\
\rotatebox[origin=c]{90}{Monopoly} & \raisebox{-.5\height}{\includegraphics[width=\PlotW]{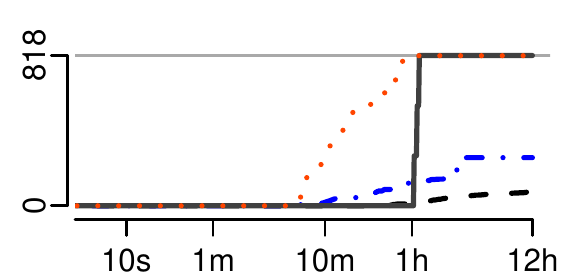}} & \raisebox{-.5\height}{\includegraphics[width=\PlotW]{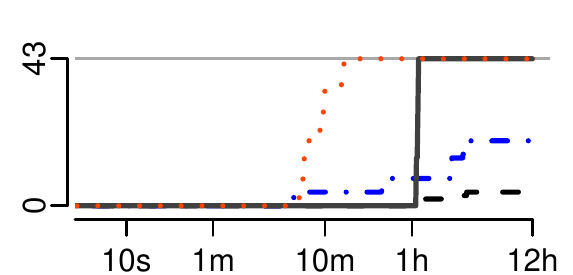}}\\
\rotatebox[origin=c]{90}{CLI} &
\raisebox{-.5\height}{\includegraphics[width=\PlotW]{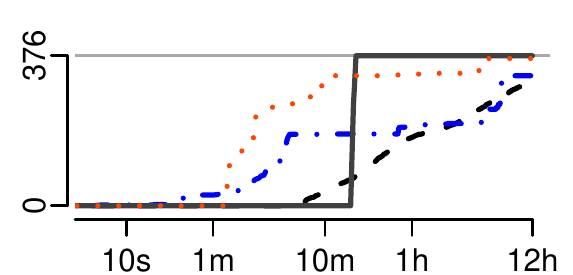}} & \raisebox{-.5\height}{\includegraphics[width=\PlotW]{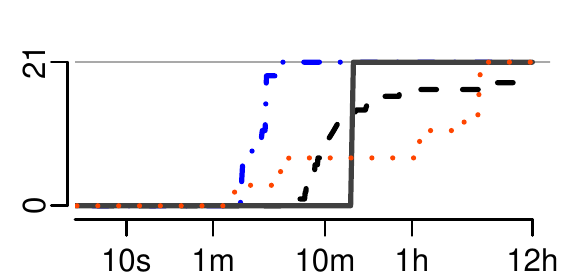}}\\
\rotatebox[origin=c]{90}{\ \ Triangle} & \raisebox{-.5\height}{\includegraphics[width=\PlotW]{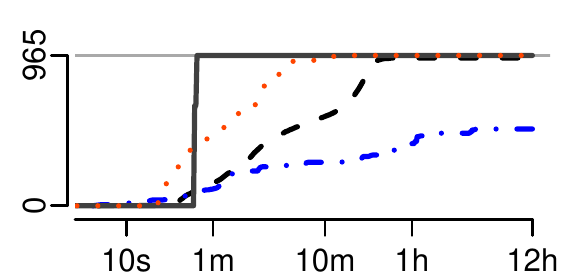}} & \raisebox{-.5\height}{\includegraphics[width=\PlotW]{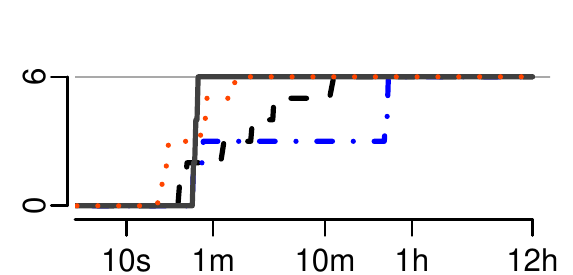}}\\
\end{tabular}
\vskip .8em
{
\small
 \includegraphics[trim=30 15 30 15,clip]{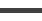} \searchvar$\quad$
 \includegraphics[trim=30 15 30 15,clip]{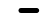}
   \searchgen\\
 \includegraphics[trim=30 15 30 15,clip]{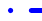}
   \searchbf$\quad$
 \includegraphics[trim=30 15 30 15,clip]{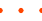}
  \searchpri\\
  {\footnotesize $\dagger$ We cap the plot for Validator since there are 13.4 billion SSHOMs; $\ddagger$ we could not enumerate all nonstrict-SSHOMs for Chess due to the difficulty of the
  SAT problem and report only those found within the time limit}
  \caption{(Strict-)SSHOMs found over time in each subject system,
  averaged over 3 executions. Note that time is plotted in log scale as most
  SSHOMs are found within the first hour.}
  \Description{SSHOM and strict-SSHOM found over time in each subject system,
  averaged over 3 executions. Note that time is plotted in log scale as most
  SSHOMs are found within the first hour.}
  	\label{fig:plots}
}
\end{figure}



\paragraph{Results}
In Table~\ref{tab:statistics}, we report the number of (strict-) SSHOMs found
with all three search strategies within the 12-hour time budget and
in Figure~\ref{fig:plots},
we plot the numbers of found (strict-)SSHOMs over time.
Note that by construction, if \searchvar\ terminates (all cases except \textit{Chess},
where solving the satisfiability problem takes considerable time), it enumerates \textit{all} SSHOMs,
thus provides an upper bound for other
search strategies---without \searchvar{} this
upper bound would not be known.

These results show clear trends:
\searchvar\ requires a relatively long time to find the first SSHOM,
because variational execution must finish executing all tests
for all combinations of
first-order mutants.
However, once variational execution finishes, 
it can enumerate \emph{all} SSHOMs very quickly
by solving
the Boolean satisfiability problem.
Variational execution takes longer with more
and longer test cases and with more first-order mutants,
but still outperforms a brute-force execution
by far,
indicating significant sharing, as found in prior analyses of
highly-configurable systems~\cite{meinickeEssential16,wongFaster18,nguyenExploring14}.

In contrast, \searchgen\ and \searchbf\
can test many candidate SSHOMs before
variational execution terminates and finds
some actual SSHOMs early, but both approaches
take a long time to find a substantial number
of SSHOMs and miss at least some SSHOMs in all
subject system within
the 12h time budget given.
In some systems with moderate numbers of first-order mutants, \searchbf\ is fairly
effective as it systematically prioritizes
pair-wise combinations which are more common
among SSHOMs than combinations of more than
two mutants, as we will discuss.

In summary, for systems where variational execution
scales, \searchvar\ can find \emph{all} SSHOMs whereas other approaches find only an
often much smaller subset within a 12h time window.
Whereas prior approaches often find their first SSHOMs
faster, \searchvar\ needs more
time upfront for variational execution but can
then enumerate SSHOMs very quickly.
To scale \searchvar{} to more realistic
programs, more engineering is needed to overcome the limitations discussed in
Sec.~\ref{subsec:limitations}. Nevertheless, \searchvar{} is valuable to the
research community as it provides a precise and efficient way of identifying
all SSHOMs.




\section{Step 2: SSHOM Characteristics}\label{sec:evaluation}

\newcommand\miniparplotW{1.8cm}
\newcommand\minibarplots[1]{
    \raisebox{-.1\height}
    {\includegraphics[width=\miniparplotW]{resources/order-#1.pdf}} &
    \raisebox{-.1\height}
    {\includegraphics[width=\miniparplotW]{resources/order-#1-strict.pdf}}
}
\newcommand\minibarplotsB[1]{
    \raisebox{-.1\height}
    {\includegraphics[width=\miniparplotW]{resources/distribution-#1.pdf}} &
    \raisebox{-.1\height}
    {\includegraphics[width=\miniparplotW]{resources/distribution-#1-strict.pdf}}
}
\newcommand\minibarplotsV[1]{
    - &
    \raisebox{-.1\height}
    {\includegraphics[width=\miniparplotW]{resources/order-#1-strict.pdf}}
}
\newcommand\minibarplotsBV[1]{
    - &
    \raisebox{-.1\height}
    {\includegraphics[width=\miniparplotW]{resources/distribution-#1-strict.pdf}}
}
\begin{table*}[t]
\centering
    \caption{Characteristics of SSHOMs and strict-SSHOMs found in our subject systems.}
    \label{tab:characteristics}
    \begin{tabular}{lccrrrrcc}
        \toprule
        & \multicolumn{2}{c}{Order} &\multicolumn{2}{c}{Equal-Fail Rule} & \multicolumn{2}{c}{N+1 Rule}& \multicolumn{2}{c}{Distribution}\\\cmidrule(r){2-3}\cmidrule(lr){4-5}\cmidrule(lr){6-7}\cmidrule(lr){8-9}
        Subject & SSHOM & strict-SSHOM & SSHOM & strict-SSHOM & SSHOM & strict-SSHOM & SSHOM & strict-SSHOM
         \\\midrule
Validator$^\dagger$ & \minibarplotsV{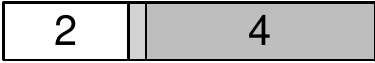}&- & 96\% & - & 99\% & \minibarplotsBV{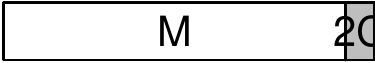}\\
Chess$^\dagger$ & \minibarplotsV{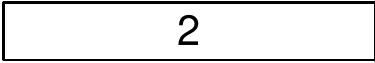}&- & 76\% & - & 38\% & \minibarplotsBV{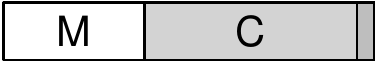}\\
Monopoly & \minibarplots{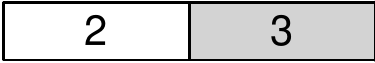}&11\% & 100\% & 99\% & 100\%& \minibarplotsB{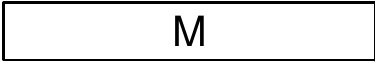}\\
Cli & \minibarplots{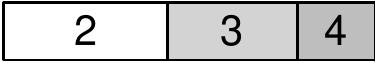}&53\% & 5\% & 98\% & 100\% & \minibarplotsB{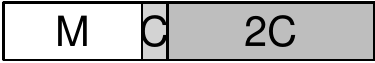}\\
Triangle &\minibarplots{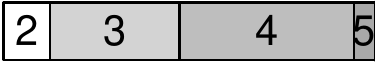}&8\% & 17\% & 98\% & 50\% & \minibarplotsB{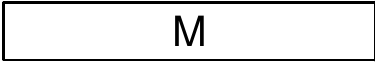}\\
        \bottomrule
    \end{tabular}
    \\[.4em]
    {\footnotesize Order counts number of constituent first-order mutants; equal-fail and N+1 rule explained in text; distribution: all constituent first-order mutants in same method (M), multiple methods in the same class (C), two classes (2C), or spread across more than two classes (*).
    \\ $\dagger$ for Validator and Chess we omit statistics, because we cannot enumerate all possible SSHOMs (too many in Validator and incomplete set in Chess)}
\end{table*}

\begin{sloppypar}
 In a second step,  we study the characteristics of (strict-) SSHOMs, with the goal to inform subsequent heuristic search strategies (Step~3) and future research in general.
 Using the \textit{complete} set derived for the subject systems in the previous step,
 rather than a (potentially biased) sample of SSHOMs,
 we can study characteristics with higher confidence.
\end{sloppypar}

We explored the dataset in an iterative exploratory fashion, focusing
primarily on characteristics that may guide future search strategies, such as
specific composition patterns and proximity of constituent first-order
mutants for the set of all higher-order mutants. \citet{kurtzAnalyzing16}
argue that mutation operators should be specialized for individual programs,
so we focus on high-level characteristics that are largely independent of
specific mutation operators to avoid overfitting. We started by randomly
sampling a large number of identified SSHOMs (among the pool of all SSHOMs).
We manually inspected the sampled SSHOMs to pose hypotheses about common
characteristics. We then operationalized the hypothesized characteristics
(i.e., develop measures to apply across all SSHOMs) to quantitatively
validate them. We repeated the process until we could not identify additional
hypotheses. Due to space constraints, we only report characteristics for
which we could quantitatively identify strong support.



\paragraph{\textbf{Mutation Order}}
SSHOMs and strict-SSHOMs are typically composed of only few first-order mutants.
Overall, over 90\,\% of all SSHOMs and strict-SSHOMs are composed of at most 4 first-order mutants,
indicating that subtle interactions are mostly caused by few first-order mutants. Although few SSHOMs
were composed of up to 6 first-order mutants (in Chess and Triangle), such cases are rare,
especially for strict-SSHOMs.
We plot the distribution of orders for both \emph{SSHOMs} and \emph{strict-SSHOMs} in Table~\ref{tab:characteristics}.

\paragraph{\textbf{Equivalent Test Failures}}
In multiple subject systems, many SSHOMs and strict-SSHOMs
are composed of first-order mutants that are killed by the
exact same set of test cases
(nonstrict-SSHOMs are often killed by the same
test cases, whereas strict-SSHOMs necessarily are
killed by fewer).
In Table~\ref{tab:characteristics}, we report how many of
the SSHOMs and strict-SSHOMs in each project
could be found when only combining first-order
mutants that are killed by the exact same
test cases, which we name \emph{Equal-Fail SSHOMs}.

\paragraph{\textbf{Containment Relationships}}
In addition,
we found a common containment pattern:
when a (strict-)SSHOM is composed of more than
two first-order mutants, it is very likely that a
subset of these first-order mutants also form
a (strict-)SSHOM.
In other words, an \emph{N+1 Rule}, combining a previously identified
(strict-)SSHOM with one further first-order mutant
is a promising strategy to identify more (strict-)SSHOMs.
In Table~\ref{tab:characteristics}, we report how many of
the \mbox{(strict-)SSHOMs} in each project
with more than two constituent first-order mutants
could be generated with such a rule.


\paragraph{\textbf{Proximity}}
Finally, for most SSHOMs, all constituent first-order mutants
are in the same class and often
even in the same method,
likely because first-order mutants with close
proximity have higher chances of data-flow
or control-flow interactions.
The effect is even more pronounced for
strict-SSHOMs. This stronger effect was previously conjectured
though not validated
\cite{jiaHigher09}.
We plot the distributions for all subject
systems in Table~\ref{tab:characteristics}.

\paragraph{\textbf{Other}}
We also explored other patterns that may inform
search heuristics, 
such as common
combinations of mutation operators (using frequent-itemset mining~\cite{agrawalMining93}),
but found no additional strong patterns. 
While we believe 
a qualitative analysis
of the mutants and their characteristics 
may reveal interesting insights about SSHOMs and whether they more closely
mirror realistic human-made faults, such analysis goes beyond
our scope of finding SSHOMs efficiently.
\looseness=-1

%% file: resources/stats.tex
Validator & 7,563 & 302 &(83\%) & 1941 &(97\%) & 1.34 * $10 ^{10}$ & 4,041 & 273 & 36,995 & 281 & 0 & 4 & 10\\
Chess & 4,754 & 847 &(84\%) & 956 &(26\%) & {3268$^\dagger$} & 
484 & 19 & 16,403 & 216 & 0 & 6 & 24\\
Monopoly & 4,173 & 99 &(89\%) & 366 &(90\%) & 818 & 81 & 349 & 817 & 43 & 4 & 15 & 43\\
Cli & 1,585 & 149 &(95\%) & 249 &(51\%) & 376 & 309 & 326 & 369 & 21 & 18 & 21 & 21\\
Triangle & 19 & 26 &(100\%) & 128 &(100\%) & 965 & 949 & 493 & 965 & 6 & 6 & 6 & 6\\\addlinespace

Ant & 108,622 & 1354 &(77\%)
& 18,280  & (92\%) & - & 1 & 0 & 44,496 & - & 0 & 0 & 61 \\
Math & 104,506 & 5177 &(79\%)
& 103,663 &(100\%) & - & 0 & 0& 390,533  & - & 0 & 0 & 2,830 \\
JFreeChart & 90,481 & 2169 &(99\%)
& 36,307 &(99\%) & - & 0 & 6 & 576,725 & - & 0 & 0 & 513 \\

%% file: sections/algorithm.tex
\section{Step 3: Characteristics-Based Prioritized Search Heuristic
\texorpdfstring{(\searchpri{})}{}}
\label{sec:algorithm}

In a final third step, we develop a new search strategy using heuristics based
on characteristics found in Step 2, which will be an incomplete, but practical alternative to our \searchvar{} strategy.


\subsection{Search Strategy}\label{subsec:priImpl}

Our new search strategy \searchpri{} avoids the overhead of
variational execution, but instead again evaluates each candidate
higher-order mutant by executing the corresponding test suite,
one candidate mutant at a time just like  \searchbf{} and \searchgen{}. 
Our key contribution is ordering how we explore candidate mutants
to steer the search toward more likely candidates. That is, instead of
a naive enumeration of all combinations (\searchbf{}) or an exploration
based on random seeds (\searchgen{}), we prioritize based on the previously identified typical characteristics
of higher-order mutants.
Since characteristics for SSHOM and strict-SSHOM do not differ strongly,
we develop only a single search strategy.

Conceptually, we calculate a penalty for every candidate higher-order
mutant and prioritize those candidates with the lowest penalty. We compute the weighted 
sum of three factors:
     \begin{equation}
     penalty =  \omega_{1} \cdot \textit{order} + \omega_{2} \cdot \textit{testDiff} - \omega_{3} \cdot \textit{isN1}
     \label{eq:penalty}
     \end{equation}
    %
First, we assign penalties based on the number of constituent first-order mutants 
(\textit{order}):
    a candidate with a higher order receives a larger penalty than a lower-order candidate,
    thus, prioritizing candidates with lower order that, as our data shows,
    are more likely to be SSHOMs.
Second, we penalize candidates constructed from first-order mutants that do not
get killed by the same test cases (\textit{testDiff}, counting the number of test cases that can kill only a subset of all constituent first order mutants), generalizing our  
    \emph{Equivalent Test Failures} insight: if all first-order mutants are killed by the 
    exact same test cases, 
    the candidate is likely to be a SSHOM, and thus gets a 0 penalty, whereas
    mutants that are killed by different test cases are less likely to form a SSHOM, 
    and thus is deferred with a higher penalty. 
Finally, we reduce the penalty of a candidate if the \emph{N+1 Rule} applies (\textit{isN1}, returning
$1$ or $0$); 
    that is, if a candidate can be constructed by adding one more first-order mutant to a known
    SSHOM, the candidate receives a boost and gets prioritized. 
By default and for our evaluation, we assign the weights    
$\omega_1=5$,
$\omega_2=1$, and
$\omega_3=15$,  based on our experience with the 
subject systems in Section~\ref{sec:evaluation}.
     

Unlike previously used genetic search strategies, where the exploration order
nondeterministically depends on random mutation and crossover in every
generation, \searchpri{} explores candidates in a deterministic order
(lexical order if two candidates have the same priority).

\subsection{Implementation}
Since we cannot enumerate and sort all possible candidate higher-order mutants
for large programs,
and even the execution of all first-order mutants may take a long time,
we devise an algorithm for \searchpri{} that identifies likely candidates in batches,
shown in Figure~\ref{fig:priorityalg}.
In each batch (configurable, by default one Java package at a time), we enumerate all 
candidate higher-order mutants up to 
a distance and order bound, then sort these candidates
by priority, and finally explore these candidates in order until
a (time) budget is reached for that batch.
Batching and bounding the search is feasible since the order and distribution
characteristics dominate the prioritization anyway and candidates beyond
those bounds would be explored only very late.
If needed batches could be revisited later with larger bounds to explore more
(less likely) candidates.

\begin{figure}[t]
    \centering
    \begin{lstlisting}[language=python, basicstyle=\ttfamily\scriptsize,frame=tb, mathescape,
    morekeywords={reachable,evaluate,enumerateCandidates,computePriorities,isSSHOM}]
def findSSHOMs(program P, mutants M, testsuite T, 
               maxOrder, maxDist, budget):
  foundSSHOMs = $\emptyset$
  # explore the program one fragment at a time
  for (batch $\leftarrow$ fragments(P)):
    # identify reachable first-order mutants in fragment
    mutants = reachable(M, batch)
    # run tests on reachable first-order mutants
    fomTestResults = for (m $\leftarrow$ mutants) evaluate(T, {m})
        
    # enumerate candidate SSHOMs up to order and distance bounds
    candidates = enumerateCandidates(mutants, maxOrder, maxDist)
    # compute priorities for each candidate
    priorities = computePriorities(candidates, fomTestResults, {})
        
    # explore candidates in decreasing priority
    while (candidates $\neq\emptyset$ $\wedge$ within budget):
      candidate = getNext(candidates, priorities)
      candidates -= candidate
      homTestResult = evaluate(T, candidate)
      if (isSSHOM(fomTestResults, homTestResult)):
        foundSSHOMs += candidate
        # update priorities based on N+1 rule
        priorities = computePriorities(candidates, fomTestResults, 
                                       foundSSHOMs)
  return foundSSHOMs
    \end{lstlisting}
    \caption{Characteristics-based prioritized search algorithm.}
    \Description{The algorithm of our characteristics-based prioritized search.}
    \label{fig:priorityalg}
\end{figure}

After batching, our algorithm identifies all first-order mutants 
defined within the given batch (function \lstinline.reachable.)
and runs the test suite for each of these first-order mutants to
identify which tests fail (function \lstinline.evaluate.).
Subsequently, the algorithm enumerates all candidates (function \lstinline.enumerateCandidates.) up to
a given order bound (by default, mutants composed of
up to 6 first-order mutants) and up to a given distance bound 
(by default, up to 4~methods spread across at most 3~classes).
Having a manageable set of candidates in the given batch,
the algorithm computes priorities (function \lstinline.computePriorities.) for all candidates 
using Equation~\ref{eq:penalty} and
then explores these candidates in order of decreasing priorities (function \lstinline.getNext.)
until either all candidates are explored or a
(time) budget has been reached in that batch
(by default, 1~hour per batch).
For each candidate, it runs the test suite and compares test results
to determine whether a (strict) SSHOM has been found (function \lstinline.isSSHOM.);
identified SSHOMs are collected and used to recompute priorities 
based on additional information for the N+1 rule.\looseness=-1

\subsection{Evaluation}\label{subsec:priEval}\label{sec:limitations2}

We evaluate how \textit{effective} our new search heuristic \searchpri{} is
at finding (strict-)SSHOMs, and 
additionally evaluate how it \textit{generalizes} and \textit{scales} to much larger systems than 
the ones used in prior studies on SSHOMs (and used in Sec.~\ref{sec:evaluation}).

\begin{figure}[t]
\centering
\begin{tabular}{rcc}
& SSHOM & Strict-SSHOM \\
\rotatebox[origin=c]{90}{Ant} & \raisebox{-.5\height}{\includegraphics[width=\PlotW]{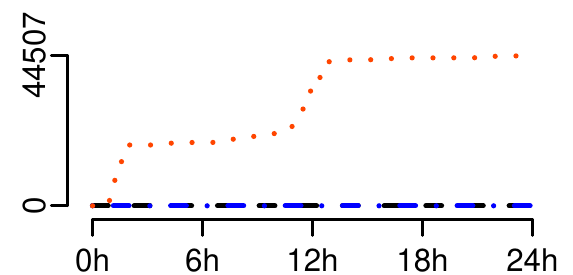}} & \raisebox{-.5\height}{\includegraphics[width=\PlotW]{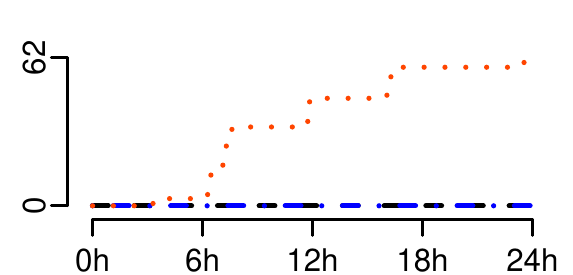}}\\
\rotatebox[origin=c]{90}{Math} & \raisebox{-.5\height}{\includegraphics[width=\PlotW]{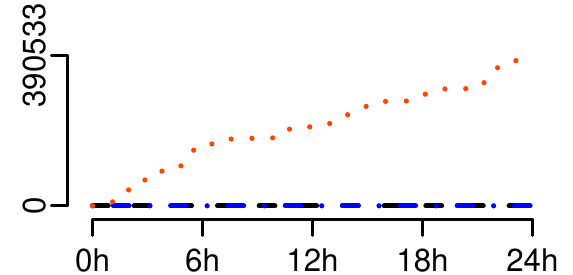}} & \raisebox{-.5\height}{\includegraphics[width=\PlotW]{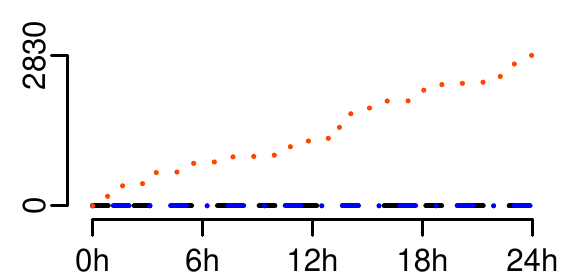}}\\
\rotatebox[origin=c]{90}{JFreeChart} & \raisebox{-.5\height}{\includegraphics[width=\PlotW]{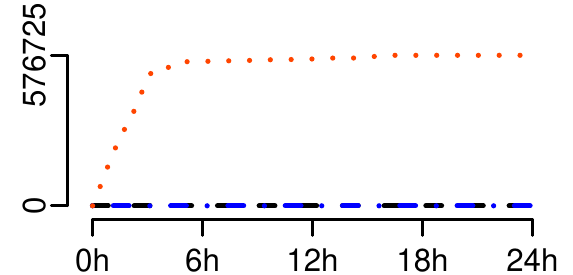}} & \raisebox{-.5\height}{\includegraphics[width=\PlotW]{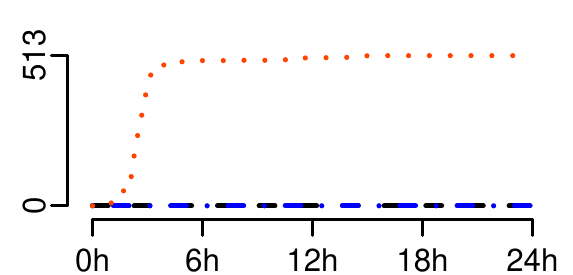}}
\end{tabular}
\vskip .8em
{
\small
 \includegraphics[trim=30 15 30 15,clip]{data/line2}\searchgen$\quad$
 \includegraphics[trim=30 15 30 15,clip]{data/line3}\searchbf$\quad$
 \includegraphics[trim=30 15 30 15,clip]{data/line4}\searchpri
  \caption{(Strict-)SSHOMs found over time,
  averaged over 3 executions. Note that time is plotted in linear scale as
  SSHOMs are found consistently over time due to batching.}
  \Description{SSHOM and strict-SSHOM found over time in each subject system,
  averaged over 3 executions. Note that time is plotted in linear scale as
  SSHOMs are found consistently over time due to batching.}
  	\label{fig:plotsnew}
}
\end{figure}

\paragraph{Subject Systems} 
We evaluate \searchpri{} both on the subjects previously
used in Section~\ref{sec:evaluation} and on a fresh set of much larger
subject systems.
The comparison against the 5~previously used subject systems allows us
to compare effectiveness against the ground truth derived from variational execution,
but the results may suffer from overfitting, as we evaluate the search
strategy on systems from which the insights that drive its design have been
derived.
\looseness=-1

Hence, we use 3~additional subjects, listed in Table~\ref{tab:statistics} (bottom),
after finishing the design of our new search strategy.
The new systems are significantly larger, allowing us to explore
the different search strategies at a much larger (and possibly more realistic)
scale.
To select the new subject systems, we collected all research
papers published in the last 5 years at ASE, FSE, and ICSE
that have the word ``mutation'' or ``mutant'' in the title. We then
selected the five largest Java systems used, discarding two 
for which we failed to reliably execute the tests. 
We did not run \searchvar{} on these systems,
but we still had to exclude some tests or mutants (reported in Table~\ref{tab:statistics}), 
due to technical issues like hard-to-terminate infinite loops. 
\looseness=-1

\paragraph{Measurements} 
We mirror our previous setup in Section~\ref{subsec:oldEval} and count the number 
of \mbox{(strict-)SSHOMs} found over time. 
We collect measurements for \searchbf{}, \searchgen{}, and 
\searchpri{}. We omit \searchvar{} for the new systems
due to engineering and scalability issues discussed in Section~\ref{subsec:limitations},
especially issues with environment barrier.
Experiments on the small subject systems were performed on the
same AWS EC2 instances (Section~\ref{subsec:oldEval}). For the new systems, we
collected measurements on Linux machines with
1.30GHz Intel i5 CPU and 16GB memory. 
When using \searchpri{}, we did not need to perform batching for the small subject systems;
we used batching for the new larger subject systems, one package
at a time, with a 1~hour budget for each package;
all other parameters were left at their defaults
(described above).
For the new subject systems, we ran each measurement for 24 hours, 
repeated \searchgen{} 3 times.

All search strategies (except \searchvar{}, not considered here) require executing the test suite
repeatedly for each candidate SSHOM. For the larger
systems, long test-execution times 
severely limit the number of mutants we can explore. To minimize
the slowdown from test execution that affects all approaches equally, 
we implement a standard regression test selection technique~\cite{papadakisMutation19} that only executes
test cases that can reach the candidate mutant
(technically, we instrument the program to record which test
reaches the location of each first-order mutant and only execute
tests that reach at least one first-order mutant of a candidate
higher-order mutant).
We apply this test optimization for all search strategies.

\paragraph{Results}
On the small subject systems, as shown in Table~\ref{tab:statistics} and Figure~\ref{fig:plots}, our new search strategy \searchpri{}
is often very effective, performing at least as well as and usually significantly outperforming both
\searchbf{} and \searchgen{} in all subjects. In a few cases, it even outperforms
\searchvar{}: In \textit{Monopoly} it finds almost all higher-order mutants before
variational execution finishes running the tests and in \textit{Chess} it finds
SSHOMs quickly, not limited by the effort to solve large satisfiability problems.

For the new and larger systems, our results shown in Table~\ref{tab:statistics} and Figure~\ref{fig:plotsnew}
show that the baseline approaches perform very poorly at this scale. Without being informed
by SSHOM characteristics the search in this vast space (e.g., 5 billion candidate combinations of mutation pairs in Math) these approaches find rarely any SSHOMs even when run for a long
time.
In contrast, \searchpri{} finds a significant number of (strict-)SSHOMs in each of these systems:
Within 24 hours it explores most batches (91\,\% of all packages) and has a reasonable 
precision~\footnote{It is difficult to fairly compare the precision of \searchpri{} with \searchvar{} and \searchbf{}, because \searchpri{} is optimized to find SSHOMs early while \searchvar{} might get better over time after a few generations. Eventually, the precision of all search-based approaches converge to almost 0 for the smaller subjects after prolonged searching. We provide more data about precision in the appendix.} for finding actual SSHOMs among the tested candidates 
($60.9\,\%$ in \textit{Math}, $29.4\,\%$ in \textit{Ant}, and $77.8\,\%$ in \textit{JFreeChart}).


We conclude that \searchpri{} is an effective search strategy that scales to large
systems and generalizes beyond systems from which the characteristics have been collected.
While we cannot assess how many SSHOMs we are missing, our strategy is effective
at finding a very large number of them in a short amount of time.




%% file: sections/relatedwork.tex
\section{Threats to Validity} 

External validity 
might be limited by the specific programs, mutation operators and test 
cases. We used subject systems from previous papers to avoid
any own sampling bias.
From most subject systems, we had to remove some tests or mutations
due to technical problems, either engineering limitations of variational
execution 
or issues
with memory leaks and infinite loops,
which might affect the results to some degree---though
we do not expect to see a systematic bias. \looseness=-1

Our study only considers three  
representative mutation operators among all possible ones~\cite{papadakisMutation19} and may not generalize to other operators.
A further analysis of the sensitivity of SSHOMs to a wide array of mutation operators
is outside the scope of this paper.

Regarding internal validity,
like other studies, our results might be affected by possible mistakes in our implementations 
or measurements and especially by we reimplemented the existing
\searchgen{} approach. To mitigate this issue, we verified that the SSHOMs found by \searchgen{} and \searchbf{}
are a strict subset of the ones found by \searchvar{}. For SSHOMs found only by our approach, we additionally verified a sample manually to ensure they are SSHOMs. 
\looseness=-1

To reduce the impact of nondeterminism in performance measurements and genetic search,
we report averages across 3 runs, as in previous work~\cite{harmanAngels14}.
Most differences are large, far exceeding the margins of error from nondeterminism or measurement noise.

As in previous work, SSHOMs should not be affected by equivalent mutants, because their specification
(Equation~\ref{eq:sshom}) explicitly requires at least one test to fail for
all combined first-order mutants. In contrast to \citet{harmanAngels14},
we do not try to establish how many of our first-order mutants are 
equivalent mutants, because we do not compute any metrics based on
the number of first-order mutants (such as `test effectiveness' in
prior studies~\cite{harmanAngels14}).

Finally, it would be possible to improve \searchbf{} and \searchgen{} by applying insights from 
our research, such as a similar batching strategy to explore one Java package at a time
and possibly also other insights from analyzing SSHOM characteristics.
When using batching (results not shown), these approaches indeed perform better on the
large subject systems but are still 
significantly outperformed by \searchpri{}.

\section{Related Work}
\label{sec:relatedwork}

In this section, we focus our discussions on higher-order mutation testing, 
and refer interested readers to a detailed survey for
recent advances in mutation testing in general~\cite{papadakisMutation19}.
\looseness=-1

\paragraph{Approaches for Finding SSHOMs}
Early work has investigated different strategies to combine first-order 
mutants into second-order mutants~\cite{mateoValidating13, kintisEvaluating10, madeyskiOvercoming14}. 
Jia and Harman extended this effort to even higher orders using heuristic 
search looking for certain kinds of valuable higher-order mutants, specifically SSHOMs.
They compare a greedy, a hill-climbing, and a genetic 
algorithm and found that genetic search produces the best results for finding 
SSHOMs~\cite{jiaConstructing08,jiaHigher09}.
Since then, higher-order mutation testing has been implemented in different mutation 
testing tools and frameworks, for different 
languages~\cite{jiaMILU08,omarHOMAJ14,kusanoCCmutator13,mahajanFinding14,leMuCheck14,tokumotoMuVM16,parsaiLittleDarwin17}, 
usually using some form of heuristic 
search~\cite{omarSubtle17,jiaConstructing08,jiaHigher09,langdonEfficient10}.
Although this
work specifically targets SSHOMs, our approach can be generalized to other types
of interesting mutants, by updating the way we encode the search as a Boolean
satisfiability problem.

\looseness=-1
Orthogonal to SSHOMs, researchers have recently investigated another
interesting type of hard-to-kill mutants called \textit{dominator mutants}~\cite{kurtzMutant14,kurtzStatic15}. This work
searches for the hardest-to-kill mutants among a set of first-order
mutants, by comparing executions with a given fixed test suite. 
Despite the heavy cost of computing dominator mutants,
they have been shown to be an effective research tool to study existing
mutation testing techniques, for example for gauging mutation test
completeness~\cite{kurtzAre16} and evaluating selective
mutation~\cite{kurtzAnalyzing16}. 
More recently, \citet{justInferring17} show that program context can be used
to approximate dominator mutants, which might also be promising for
future search strategies for SSHOMs.



\paragraph{Characteristics of SSHOMs}
Existing studies on SSHOMs mostly concern the quantity 
of SSHOMs and difficulty of finding
them~\cite{harmanAngels14,jiaHigher09,jiaConstructing08,jiaMILU08,langdonEfficient10}. 
For example, \citet{harmanAngels14} discussed how SSHOMs relate
to their constituent first-order mutants, but their discussion focuses mainly on test effectiveness and 
efficiency.
\citet{jiaHigher09} discussed characteristics of a single SSHOM in the Triangle program (also used in our study), but did not explore SSHOM characteristics further.
In our work, we can find a complete set of SSHOMs, which provides us more data 
to study what they look like. 

\paragraph{Variational Execution}
Variational execution was originally developed for information-flow analysis~\cite{austinMultiple12} and configuration testing~\cite{nguyenExploring14,KKB:ISSRE12}.
In a new-idea paper, \citet{wongTesting18} recently suggested that variational execution may
have additional application scenarios, suggesting mutation testing as explored in
Step 1 of this paper as one promising direction.

\looseness=-1
With regard to using variational execution for mutation testing,
\citet{devroeyFeatured16} are conceptually closest to our work in that they
pursue a complete exploration strategy with similarities to lazy configuration
exploration in SPLat~\cite{KMS+13,meinickeEssential16}.
However, they explore only traces in state machines without any
joining and thus forgo much
possible sharing. Their analysis does not distinguish first-order from
higher-order mutants and does not identify or analyze SSHOM.
Several other researchers have also used advanced dynamic analyses
to speed up the execution of tests in traditional mutation testing
(one mutation at a time), looking for possible redundancies and joins
\cite{wangFaster17,JH:TSE11,justEfficient14}. Since our main goal of using 
variational execution is to explore interactions of first-order mutants rather than speed up mutation analysis, we did not perform a performance comparison.

%% file: sections/conclusion.tex
\section{Conclusions}

To efficiently find SSHOMs, we proceed in three steps.
First, we use variational execution to find \textit{all} SSHOMs in small to medium-sized programs. 
Second, we analyze basic characteristics of the identified SSHOMs. 
Finally, we derive a new prioritized search strategy based on the characteristics. 
The prioritized search scales
to large systems and is effective (albeit not complete) at finding SSHOMs
and outperforms the existing state-of-the-art strategy by far.
We hope that the insights and search strategies from this work can support future work 
in mutation testing.

%% file: sections/appendix.tex

\newcommand\searchideal{$\texttt{search}_\textit{ideal}$}
\section{Discussion on Precision}
\noindent
In this section we discuss the precision of our new prioritization-based approach (\searchpri{}) compared to the genetic algorithm (\searchgen{}) and brute-force (\searchbf{}).
In general, it is difficult to fairly compare the approaches as they excel in different stages of the search, and all search-based approaches converge to a low precision after running the algorithm for a long time.\\[-0.7em]

\paragraph{Different Stages of the Search} While \searchpri{} is designed to find SSHOMs early, \searchgen{} initially starts with a random set of candidates, which means it is essentially a random search in the early stage. 
The genetic search could get better over time when the fitness function becomes useful in guiding the search, but it is difficult to predict due to the stochastic nature of genetic algorithm. Also, the initial precision of \searchgen{} and \searchbf{} can be influenced by the order of first-order mutants to combine.
For example, if \emph{mut\_47} is contained in many SSHOMs then the approaches might be more precise if \emph{mut\_47} is selected early instead of later.\\[-0.7em]


\paragraph{Long Running Search}
All three search-based approaches eventually tend to find significantly fewer solutions (e.g., get stuck in a local optimum). Thus, the precision gets lower, even close to 0\% if the number of SSHOMs is small comparing to the search space.
Thus, measuring the overall precision of the approaches after, for example, 12 hours reveals limited insights.\\[-0.7em]

\noindent To give an intuition of precision, we compare the studied approaches with 
an \emph{ideal} approach (\searchideal{}) that generates SSHOMs with perfect precision. 
The execution time of \searchideal{} is based on the average execution time of the test suite.
That is, if the test suite takes one second to execute then the approach would take 100 seconds to find and evaluate 100 SSHOMs.

\begin{figure}[tb]
\begin{tabular}{rcc}
& Linear view & Focused view \\
    \rotatebox[origin=c]{90}{Triangle} & \raisebox{-.5\height}{\includegraphics[width=\PlotW]{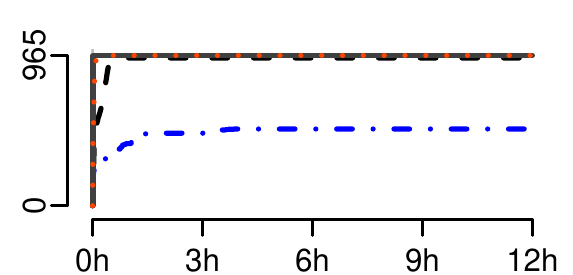}} & 
    \raisebox{-.5\height}{\includegraphics[width=\PlotW]{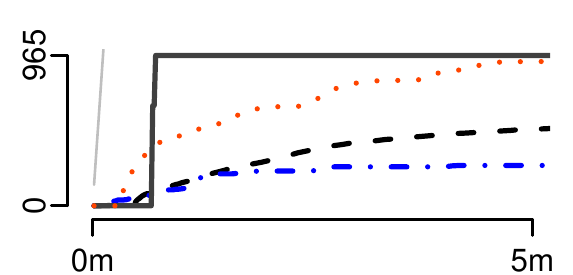}}\\
    \rotatebox[origin=c]{90}{Cli} & \raisebox{-.5\height}{\includegraphics[width=\PlotW]{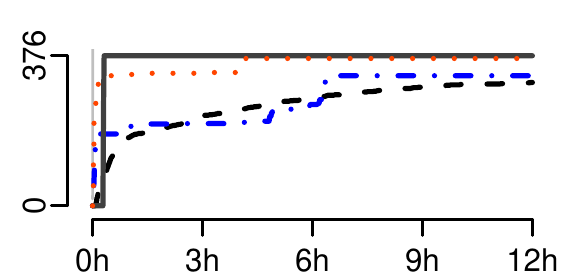}}& \raisebox{-.5\height}{\includegraphics[width=\PlotW]{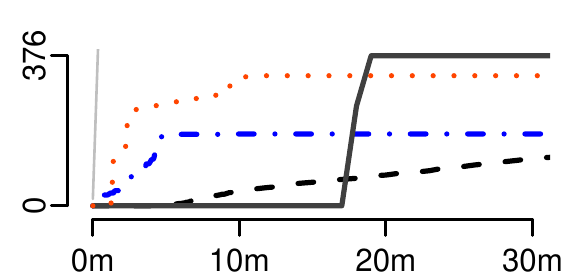}}\\
    \rotatebox[origin=c]{90}{Monopoly} & \raisebox{-.5\height}{\includegraphics[width=\PlotW]{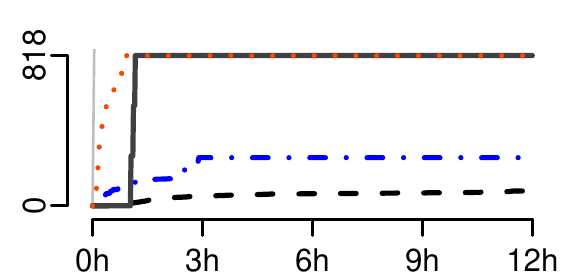}}& \raisebox{-.5\height}{\includegraphics[width=\PlotW]{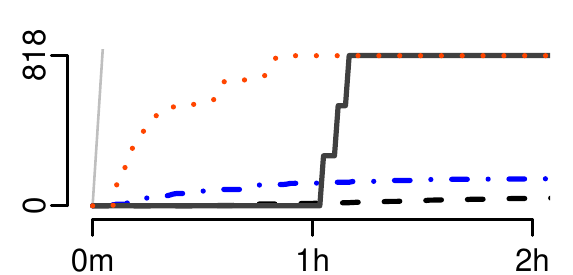}}\\
    \rotatebox[origin=c]{90}{Chess} & \raisebox{-.5\height}{\includegraphics[width=\PlotW]{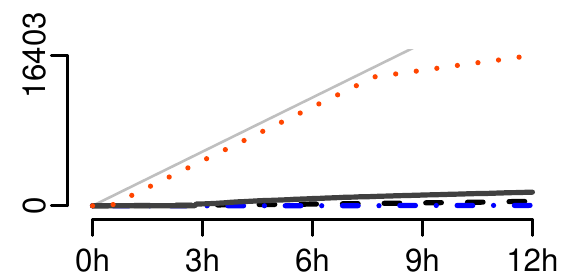}}& \raisebox{-.5\height}{\includegraphics[width=\PlotW]{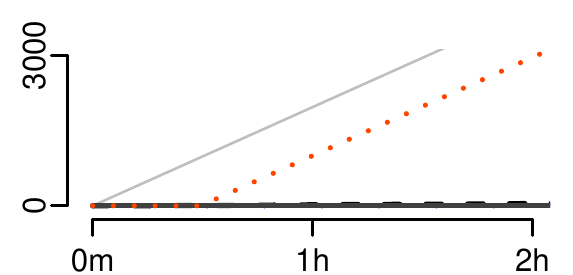}}\\
    \rotatebox[origin=c]{90}{Validator} & \raisebox{-.5\height}{\includegraphics[width=\PlotW]{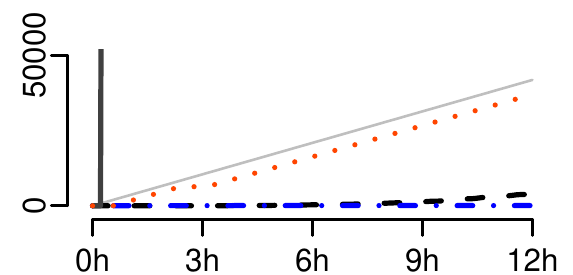}}& \raisebox{-.5\height}{\includegraphics[width=\PlotW]{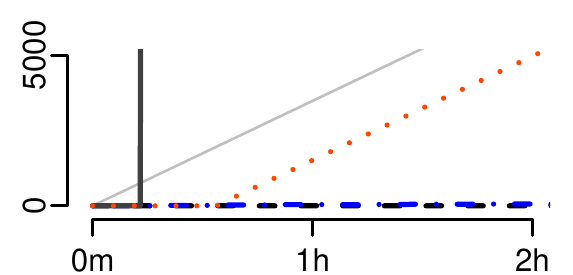}}
    \end{tabular}
    \vskip .8em
{
\small
 \includegraphics[trim=30 15 30 15,clip]{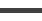} \searchvar$\quad$
 \includegraphics[trim=30 15 30 15,clip]{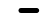}
   \searchgen
   \\
 \includegraphics[trim=30 15 30 15,clip]{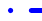}
   \searchbf$\quad$
 \includegraphics[trim=30 15 30 15,clip]{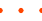}
  \searchpri$\quad$
  \includegraphics[trim=30 15 30 15,clip]{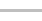}
  \searchideal
    }
\caption{Performance of the approaches for finding SSHOMs, plotted on a linear scale. The \searchideal{} line depicts the performance of an ideal approach, as a reference to gauge the precision of other approaches over time.}
    \label{fig:performance}
\end{figure}

We show the results in Figure~\ref{fig:performance}.
The left-hand side shows the progress of finding SSHOMs within the 12 hour budget, plotted in \emph{linear scale} to give an intuitive overview of progress.
To illustrate and compare precision, we show a \emph{focused view} that focuses on the beginning of the search.
In the plots, the steepness of the curves is the precision at any given point in time. 
The steepness of \searchideal{} illustrates the maximum possible precision that search-based approaches can achieve.
In general, we can see that \searchpri{} has a high precision, especially in the very beginning of the search. 
When searching for more difficult SSHOMs, the precision gets lower. 
Especially, in Chess and Validator, we can see that the lines for \searchideal{} and \searchpri{} are almost parallel, showing that the precision is close to ideal.
Note that the shift of the lines comes from the initial effort of \searchpri{} for evaluating all first-order mutants and for generating the initial set of candidate SSHOMs.

Regarding \searchbf{}, it appears relatively efficient when searching for second-order SSHOMs in Triangle, Cli, and Monopoly.
When searching for higher orders than two, the precision drops drastically for \searchbf{}.
As discussed, the initial precision of \searchgen{} is close to a random approach, but we can see that the precision might become better over time (see, for example, Triangle and Cli).

